\documentclass[%
 reprint,
 amsmath,amssymb,
 aps,
]{revtex4-1}

\usepackage{graphicx}
\usepackage{dcolumn}
\usepackage{bm}

\begin{document}


\title{Muon capture reaction on $^{100}$Mo to study nuclear responses for double beta decays and astro-neutrinos}
\author{I.H.Hashim$^{1,2}$,Ejiri$^{2}$, T.Shima$^{2}$,A.Sato$^{3}$,Y.Kuno$^{3}$,N.Kawamura$^{4}$, S.Miyake$^{4}$, K.Ninomiya$^{4}$}
\affiliation{$^{1}$Department of Physics, Faculty of Science, Universiti Teknologi Malaysia, Johor Bahru, Malaysia.}
\altaffiliation[Also at ]{Department of Physics, Faculty of Science, Universiti Teknologi Malaysia, 81310 Skudai, Johor Bahru, Malaysia.}
\email{izyan@utm.my}
\affiliation{$^{2}$Research Center for Nuclear Physics, Osaka University, Osaka, Japan.}
\affiliation{$^{3}$Department of Physics, Osaka University, 
Osaka, Japan.}
\affiliation{$^{4}$Material Life Science Facility, J-PARC, 
Tokai, Japan.}%

\date{\today}

\begin{abstract}
The negative-muon capture reaction (MCR) on the enriched $^{100}$Mo isotope was studied for the first time to investigate neutrino nuclear responses for neutrino-less double beta decays and supernova neutrino nuclear interactions. MCR on $^{100}$Mo proceeds mainly as $^{100}$Mo($\mu$,xn)$^{100-x}$Nb with $x$ being the number of neutrons emitted from MCR. The Nb isotope mass distribution was obtained by measuring delayed $\gamma$-rays from radioactive $^{100-x}$Nb. By using the neutron emission model after MCR, the neutrino response (the strength distribution) for MCR was derived. Giant resonance (GR)-like distribution at the peak energy around 11-14 MeV, suggests concentration of the MCR strength at the muon capture GR region. 
\begin{description}
\item[DOI]                  
\item[PACS number(s)]23.40.-s, 13.35.Hb, 14.60.St.
\end{description}
\end{abstract}

\pacs{Valid PACS appear here}
\maketitle


\section{\label{sec:level1}Introduction}

Neutrino nuclear responses (square of nuclear matrix element, NME) are crucial for neutrino studies in nuclei. Neutrino properties such as the Majorana nature, the absolute mass scale and others beyond the standard electro-weak model are studied by investigating neutrino-less double beta decays (DBD) \cite{ref1,ref2,ref3,ref4,ref5}. Here the $\nu$ responses for DBD ($\beta^-$ and $\beta^+$ responses) are necessary to get the neutrino properties beyond the standard model. The DBD matrix element is expressed in terms of the product of $\beta^{-}$ and $\beta^{+}$ matrix elements of $M(\beta^{-}$) and $M(\beta^{+}$). Astro-neutrino nucleo-syntheses and astro-neutrino nuclear reactions  are studied by investigating astro neutrino nuclear interactions. Then one needs nuclear responses for astro neutrinos and astro anti-neutrinos \cite{ref1,ref2}. So far, the single $\beta ^-$ matrix element $M(\beta ^-)$ and the neutrino response have been extensively studied by charge exchange reactions \cite{ref1,ref5}.

The present report aims to show that negative muon ($\mu$) capture reaction (MCR) is used to study the neutrino nuclear responses relevant to the $\beta^+$ response involved in DBD and anti-neutrino response associated with astro anti-neutrino reactions. The MCR response for $^{100}$Mo shows a giant resonance (GR) distribution at the peak energy around 11-14 MeV.

\begin{figure}
\centering
\includegraphics[scale=0.4]{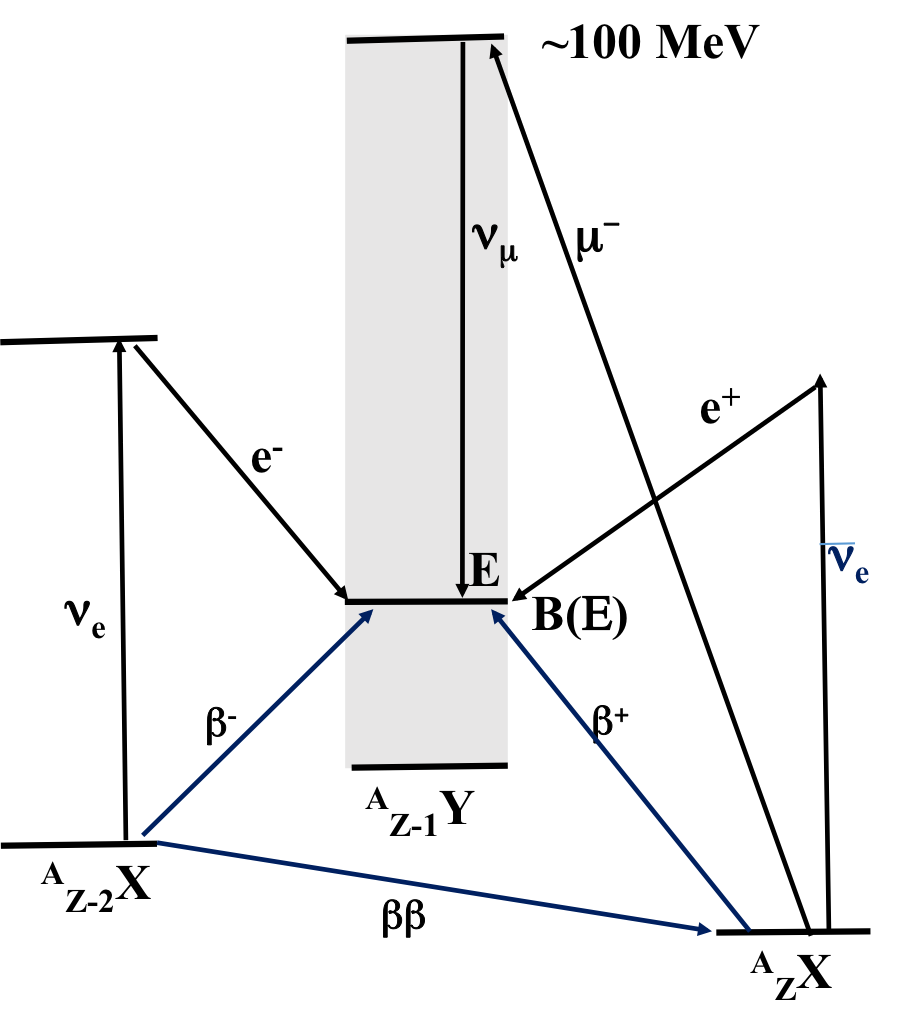}
\caption{\label{figure1}Reaction and decay schemes for astro neutrinos and double beta decays.}
\end{figure}

MCR is a kind of muon charge exchange reaction (MCER) via the weak boson W$^{\pm}$, where muon becomes muon neutrino and  proton in the target nucleus transform into neutron. The response represents the square of the nuclear matrix element $M(\mu)$. The reaction and the NME are expressed as 
\begin{equation}
\mu + ^{A}_{Z}X \rightarrow ^A_{Z-1}Y + \nu_{\mu}, ~~~ M(\mu),
\end{equation}
where $^A_ZX$ with $A$ and $Z$ being the mass number and the atomic number of the target nucleus and  $^A_{Z-1}Y$ is the nucleus after MCR. The corresponding astro anti-neutrino reaction and the NME are given as 
\begin{equation}
\bar {\nu}_e + ^{A}_{Z}X \rightarrow ^A_{Z-1}Y + e^+, ~~~ M(\bar{\nu}).
\end{equation}
DBD via the light Majorana $\nu$ exchange process with the $\beta^+$ and $\beta^-$ responses is written as the neutrino emission and re-absorption process as
\begin{equation}
^A_Z X\rightarrow ^A_{Z-1}Y + \nu_{e} + e^+, ~~~ M(\beta ^+).
\end{equation}
\begin{equation}
^A_{Z-2}X\rightarrow ^A_{Z-1}Y + \bar {\nu}_e  + e^-, ~~~ M(\beta ^-),
\end{equation}
where $^A_{Z-1}$Y is the intermediate nucleus. Note that the $\mu, \bar{\nu}$ and $\beta ^+$ NMEs are associated 
with $\tau ^+$ p$\rightarrow$n transition, while $\nu$ and $\beta ^-$ NMEs are with $\tau ^-$  n$\rightarrow$ p one. 
These reaction and decay schemes are shown in Fig. \ref{figure1}. Here the DBD NME $M(\beta \beta)$ is given by the sum of NMEs $M_i(\beta \beta )$ over all relevant states in the intermediate nucleus $^A_{Z-1}Y$. The NMEs $M_i(\beta \beta)$ are associated indirectly with the single $\beta$ NMEs of $M_i(\beta^+)$ and $M_i(\beta ^-)$ via the intermediate state i. Thus information of $M(\beta ^{\pm})$ is used to help evaluate 
$M(\beta \beta )$.

Theoretical calculations of NMEs for neutrino nuclear responses are very hard since they are very sensitive to nuclear correlations, nuclear medium effects and nuclear models, as discussed in review articles \cite{ref4,ref5}. Thus experimental studies of them are very valuable. Experimental studies of the neutrino nuclear responses for DBDs and astro neutrinos are described in review articles \cite{ref1,ref2,ref3,ref4,ref5}. 

The NMEs $M(\nu)$ for astro neutrino and the $M(\beta ^-)$ for DBD (the left hand side of Fig. \ref{figure1}) have been studied extensively by using high energy-resolution ($^{3}$He,t) experiments at RCNP \cite{ref1,ref4,ref5}. On the other hand, there are no appropriate high energy-resolution nuclear probes for the astro anti-neutrino $M(\bar{\nu})$ and DBD $M(\beta^{+})$ (the right hand side of Fig.\ref{figure1}).

\begin{figure}
\centering
\includegraphics[scale=0.5]{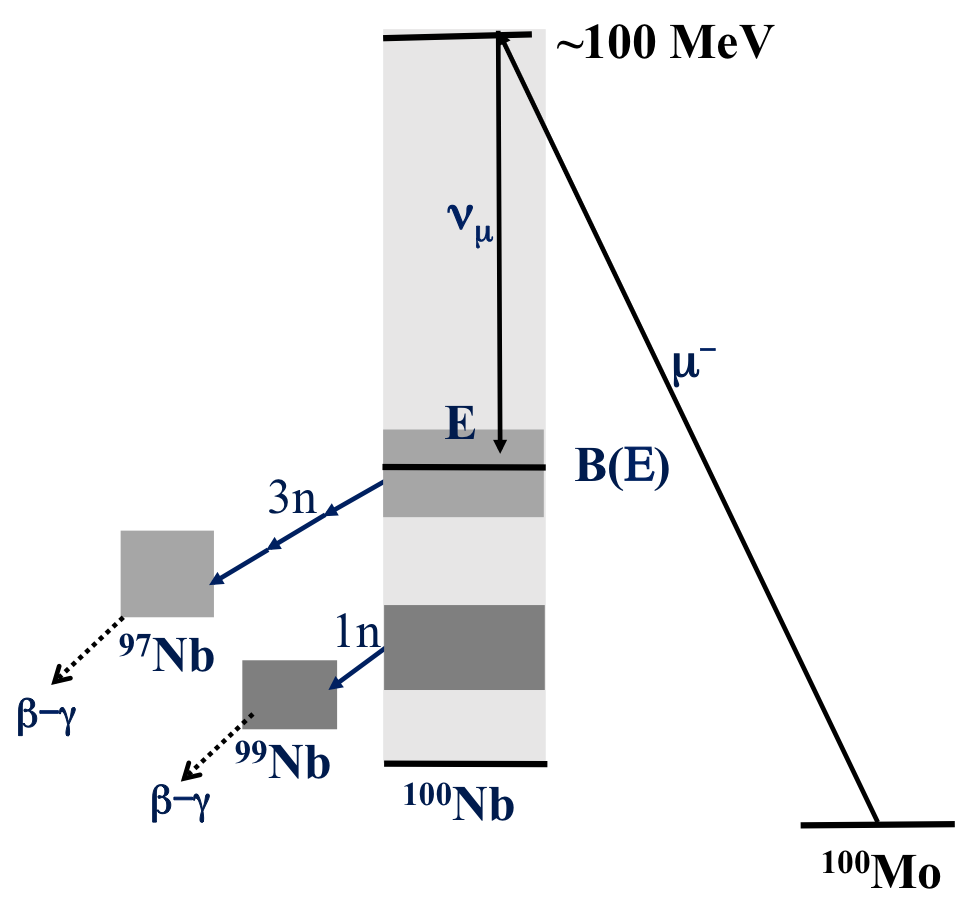}
\caption{\label{figure2}Reaction and decay schemes for MCR on $^{100}$Mo. Highly excited states in $^{100}$Nb around 30 MeV de-excite by emitting 3 neutrons, populating finally the $^{97}$Nb RI, while those around 12 MeV de-excite by emitting 1 neutron, populating finally the $^{99}$Nb. The residual nuclei are identified by observing $\gamma$-rays characteristic of the RIs.}
\end{figure}

The present MCR provides useful information relevant to $M(\bar{\nu})$ and $M(\beta^{+})$. Unique features of MCR is the large energy and momentum regions of $E$=0-50 MeV and $p$= 20-100 MeV/c, which are the regions involved in neutrino-less DBD neutrinos and medium energy supernova neutrinos. 

In MCR, a well-bound proton in the target nucleus $^A_ZX$ is shifted up to a vacant neutron shell, and one gets 
mostly the excited nucleus $^A_{Z-1}$Y$^* $ with the excitation energy $E$. If particle bound, it decays by emitting $\gamma$-rays to the ground state of $^A_{Z-1}$Y. On the other hand, if $^A_{Z-1}$Y$^* $ is unbound, it de-excites by emitting a number ($x$) of neutrons in case of medium heavy nucleus since proton emission is suppressed much by the Coulomb barrier. 

Finally one residual nucleus  $^{A-x}_{Z-1}Y$ is obtained. If it is radioactive, we identify it by measuring characteristic $\gamma $-rays of the residual nucleus. The number of the 
emitted neutrons reflects the excitation energy $E$, larger $x$ corresponds to the higher $E$ region. Thus one can derive the MCR response, i.e. the strength distribution, as a function 
of $E$, from the mass ($A-x$) distribution of the residual isotopes, as suggested in 1972 \cite{ref6}, and also in 2001's \cite{ref2,ref7,ref8}. The MCER and the neutron emission schemes are illustrated in Fig.\ref{figure2}.

So far nuclear MCR $\gamma$-rays were measured to study the nuclear reaction mechanisms \cite{ref9,ref10}. Prompt $\gamma$-rays from bound states excited by $\mu$ capture $^{A}$X($\mu$,$\gamma$)$^{A}$Y reactions were investigated to 
study $\beta^{+}$ responses for low lying bound states \cite{ref11}. Here it is hard to extract the $\beta^{+}$ 
strengths to individual states because the states are populated not only directly by MCR but also indirectly from so many higher states populated by MCR via $\gamma$ transitions. 

In the present work, we focus on gross structure of the MCER strength distribution by measuring the mass distribution of residual isotope (RI) $^{A-x}Y$. The RI yield is well obtained 
by measuring the yield of $\gamma$-rays from $^{A-x}Y$ and the known $\gamma$-branching ratio. A neutron emission code was 
developed to link the mass number $A-x$ of the residual nucleus to the initial excitation energy $E$ of $^A_{Z-1}Y^*$.

\begin{figure}
\centering
\includegraphics[width=85.0mm]{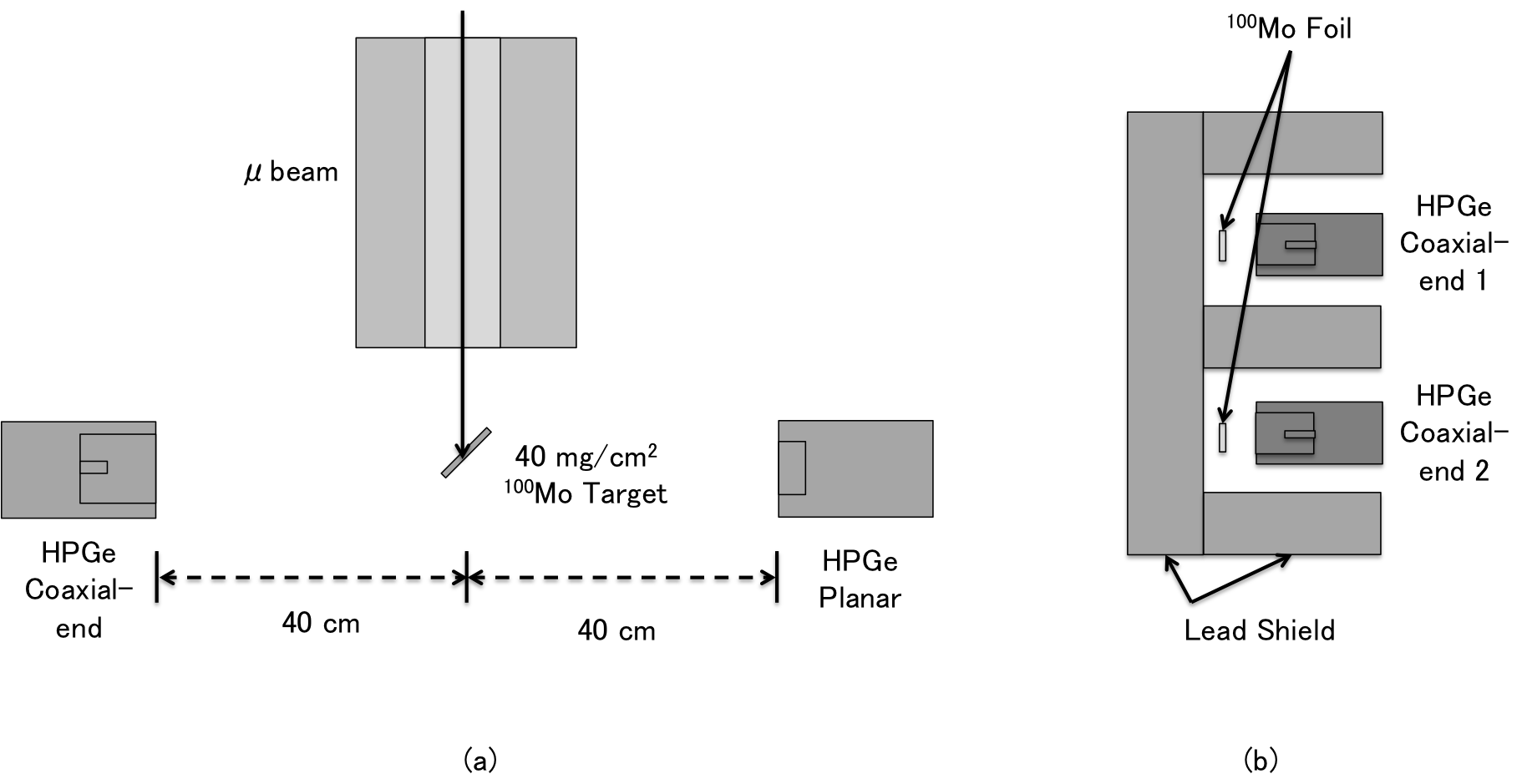}
\caption{\label{figure3}Schematic plane view of (a) the target and the coaxial-type Ge detectors for prompt X and prompt/short-lived $\gamma$-rays and (b) the Ge detectors for the off-line measurements for the long-lived RIs.}
\end{figure}

\section{\label{sec:level2}Muon Irradiation Experiment}

The present MCR experiment is made on $^{100}_{42}$Mo as a typical medium heavy nucleus for astro-physics and DBD interests \cite{ref12,ref13}. The target used is a thick $^{100}$Mo with 40 mg/cm$^2$. Low energy $\mu^-$ beams with p = 28 MeV/c from the D-beam line at J-PARC MLF were used to irradiate the target for 7 hours. Here most muons were stopped and captured into the target nucleus to form excited states in $^{100}_{41}$Nb. Then $\gamma$-rays from short-lived Nb isotopes with $A$=100 and 99 were measured on-line using planar and coaxial-end type HPGe detectors as shown in the left hand side of Fig.\ref{figure3}, while $\gamma$-rays from long-lived Nb isotopes and other isomers were measured off-line at a separate room by using 2 coaxial Ge detectors as shown in the right hand side of Fig.\ref{figure3}. 

\section{\label{sec:level3}Gamma rays from MCR on enriched Mo and Nb mass distribution}

\begin{figure}[htb!]
\centering
\includegraphics[scale=0.35]{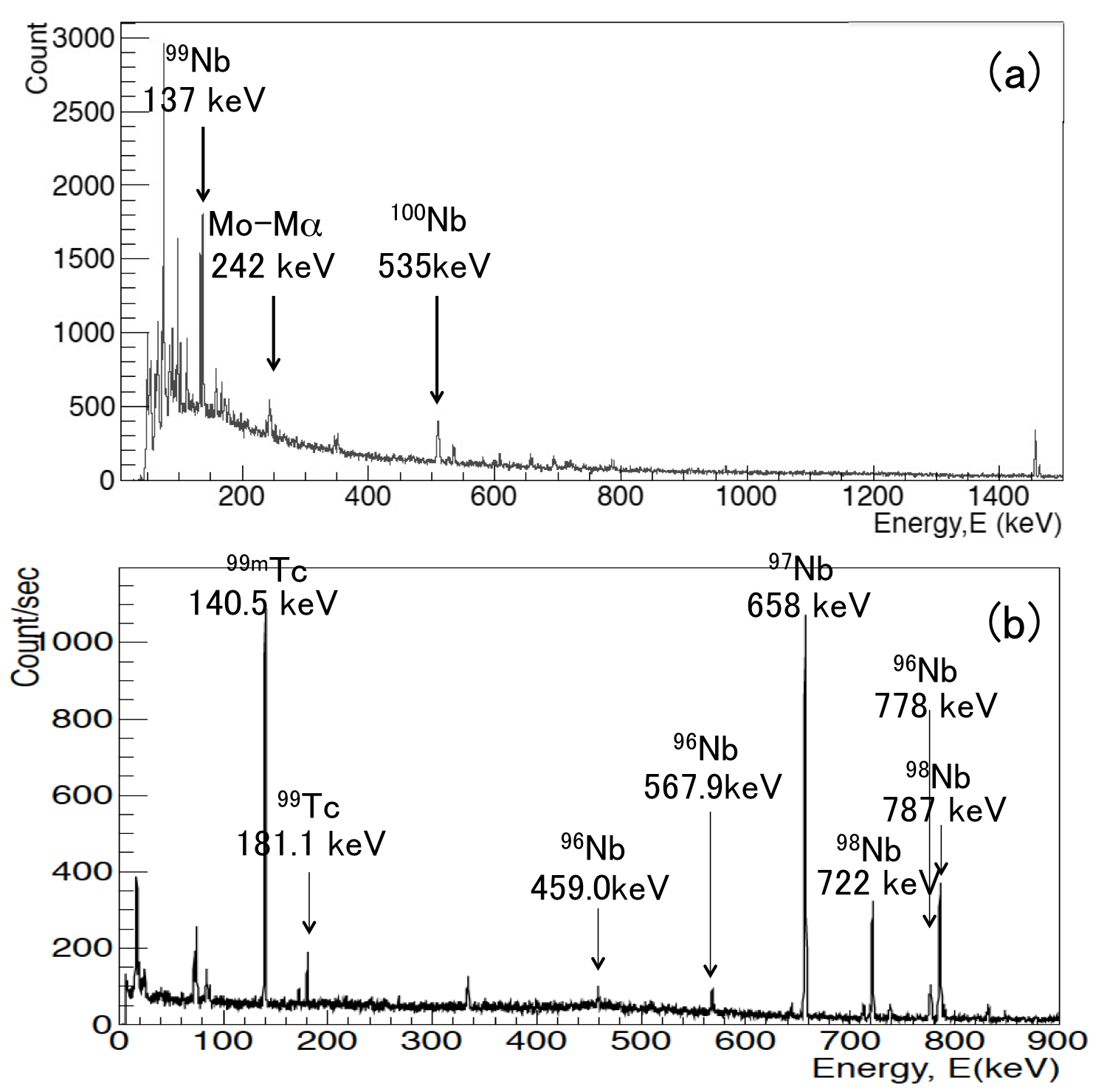}
\caption{\label{figure4} Gamma ray spectra from MCR on $^{100}$Mo. (a) On-line spectrum for the prompt $\gamma $-rays and delayed ones from  short-lived RIs. (b) Off-line spectrum for the delayed $\gamma$-rays from long-lived RIs.} 
\end{figure}

The prominent $\gamma $-ray peaks from short-lived and long-lived Nb isotopes were clearly observed as shown in Fig.\ref{figure4}. The typical $\gamma $-rays from the MCR products are listed in Table\ref{table1}. The 535.0 keV $\gamma$-ray from the short-lived $^{100}$Nb and 137 keV $\gamma$-ray from the 
short-lived  $^{99}$Nb were measured by the on-line Ge detector setting, while others from the long-lived  $^{99}$Mo, $^{99}Tc^{m}$, $^{98}$Nb, $^{97}$Nb, $^{96}$Nb, and $^{95}$Nb 
were measured by the off-line Ge detector setting. Among them, the 66 hr $^{99}$Mo is the $\beta ^-$ decay product from $^{99}$Nb and the 6 hr $^{99}$Tc$^m$ is the isomeric state produced by the $\beta ^-$ decay from $^{99}$Nb. 

\begin{figure}[htb!]
\centering
\includegraphics[scale=0.3]{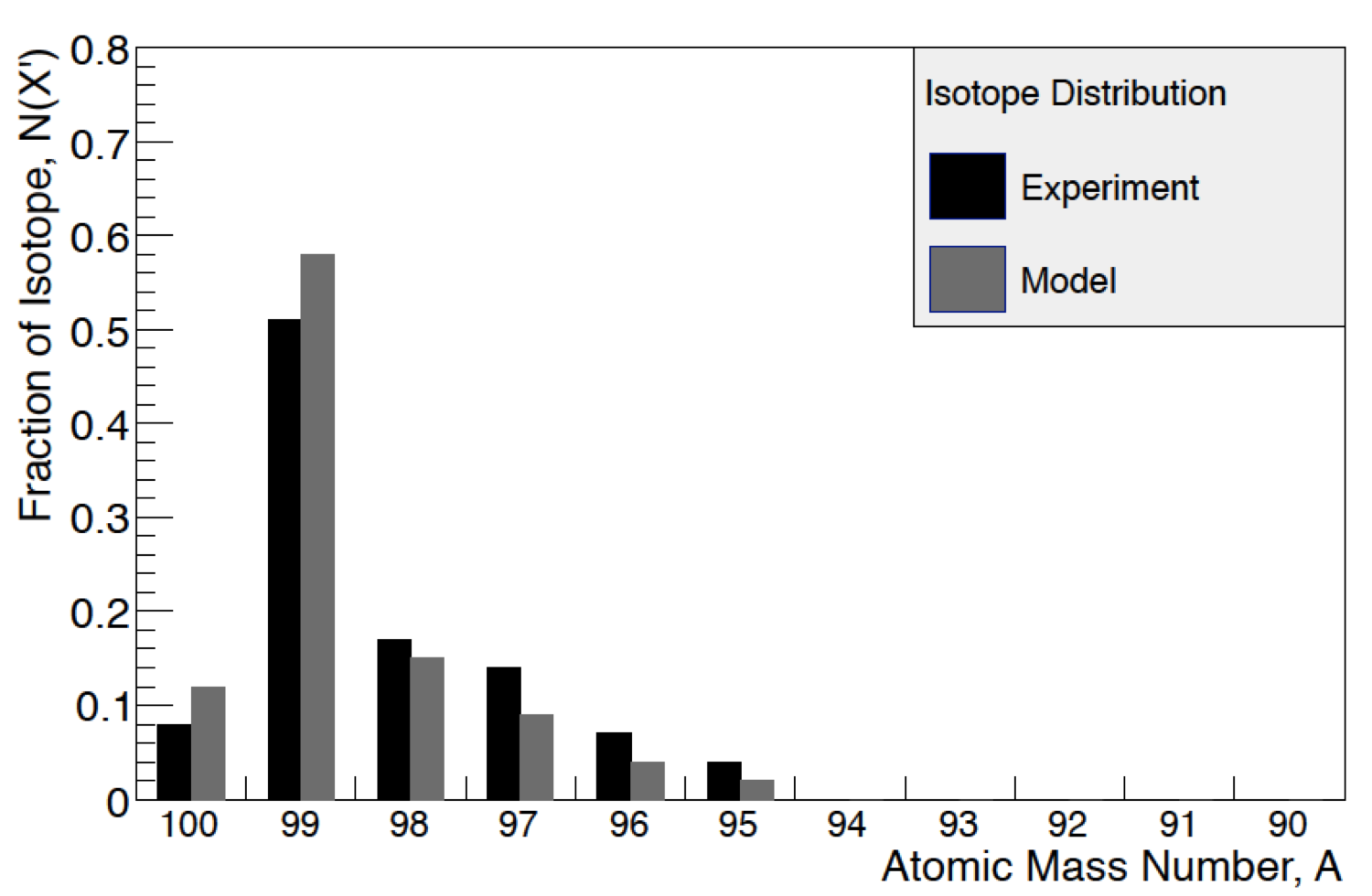}
\caption{\label{figure5} Nb RI mass distribution for MCRs on $^{100}$Mo. The black and light grey histograms are the experimental and calculated yields.}  
\end{figure}

The number of the Nb RIs $^{100-x}$Nb produced by MCR on $^{100}$Mo was evaluated from the observed $\gamma$-ray yields corrected for the Ge detector efficiency, the $\gamma $-ray branching ratio, and their decays during the muon irradiation and the $\gamma$-ray measurement. The obtained RI mass ($A-x$) distribution is shown in Fig.\ref{figure5}. The $^{100}$Nb yield at $x$=0 is quite small, but jumps up drastically at $x$=1, and decreases gradually as $x$ increases down to the mass $A$=95 and $x$=5. This is similar to the distributions in other target nuclei \cite{ref7,ref9,ref10,ref17}.

\begin{table}
\centering
\caption{\label{table1}Nuclear isotopes observed by MCR ($\mu$, xn) reactions. Half lives are given by s: second, m: minutes,  h: hour or d: day.}
\begin{tabular}{p{2cm} p{2cm} p{3.5cm} p{2cm}} 
\hline
Isotope & lifetime & Energy(keV)  \\ 
\hline
 $^{100}$Nb &2.99 s & 535.0  \\ 
 $^{99}$Nb &2.6 m& 137.0  \\
 $^{99}$Mo &66.0 & 181.1, 739.5  \\
 $^{99}$Tc & 6 h & 140.5  \\
 $^{98}$Nb &1.23 h & 722.0, 787.0  \\
 $^{97}$Nb &0.85 h & 657.9  \\
 $^{96}$Nb &23.3 h & 459.0, 568.7, 778.0  \\
 $^{95}$Nb &34.9 d & 765.0 \\
 $^{94}$Nb &51.8 m & 75.5, 366.9, 891.7 \\
 $^{93}$Nb &6.85 h & 2424.9 \\
\hline
\end{tabular}
\end{table}

Let us evaluate the RI mass distribution on the basis of the MCR strength distribution and the statistical neutron emission model. MCR excitations are expressed in terms of the vector 
excitations with the spin transfers of $\Delta J^{\pi} = 0^+, 1^-, 2^+$ and the axial-vector ones with $\Delta J^{\pi} = 1^+, 2^-$. Among them the 0$^+$ Fermi and the 1$^+$ GT excitations 
are reduced much since the 0$\hbar \omega$ Fermi and GT excitations for the $\beta ^+$ and the anti-neutrino responses are blocked by the neutron excess in medium heavy nuclei of the 
present interest. The 1$^-$ excitation with the 1$\hbar \omega$ jump may show the giant resonance (GR) like the E1 GR in case of the photon capture reaction (PCR). 

The vector 2$^+$ and the axial-vector 2$^-$ excitations 
may show a broad GR-like distributions as the 2$\hbar \omega $ and spin dipole GRs. Accordingly, we assume MCR strength distribution of $B(\mu,E)$ given by the sum of the two GR strengths of $B_1(\mu,E) $ and $B_2(\mu,E)$, 
\begin{equation}
B(\mu,E)=B_1(\mu,E) + B_2(\mu,E),  
\end{equation}
\begin{equation}
B_i(\mu,E)=\frac{B_i(\mu)}{(E-E_{Gi})^2 + (\Gamma_{i} /2)^2},
\end{equation}
where $E_{Gi}$ and $\Gamma_i$ with $i$=1 and 2 are the GR energy and the width for the $i$th GR, and the constant $B_i(\mu)$ is expressed as $B_i(\mu)=\sigma_i \Gamma _i/(2\pi)$  with 
$\sigma_i$ being the total strength integrated over the excitation energy. 

Excited states $^A_{Z-1}$Y$^*$ populated by MCR de-excite mostly by emitting neutrons at the pre-equilibrium (PEQ) and equilibrium (EQ) stages \cite{ref18}. If the states are neutron 
unbound, and do by emitting $\gamma$-rays to the ground state if they are bound. Here we ignore proton emission which is 
prohibited by the Coulomb barrier in case of the medium and heavy nuclei. The energy spectrum of the first neutron E$_n^1$ is given by \cite{ref14}.
\begingroup\small
\begin{equation}
S(E_n^1) = k [E_n^1exp (-\frac{E_n^1}{T_{EQ}(E)} + p E_n^1exp (-\frac{E_n^1}{T_{PEQ}(E)}) ],
\end{equation}
\endgroup
where $E_n^1$ is the first neutron kinetic energy, $T_{EQ}(E)$ and $T_{PEQ}(E)$ are the EQ and PEQ nuclear temperatures and $p$ is the fraction of the PEQ neutron emission. The neutron emission from the EQ stage is a kind of neutron evaporation from thermal equilibrium phase. 

The EQ temperature is expressed as $T_{EQ}(E)=\surd(E/a)$ with $a$ being the level density parameter \cite{ref14}. The parameter $a$ is expressed as $a=A$/8 MeV for the nucleus with 
mass number $A$. $T_{PEQ}(E)$ is given by $b\times T_{EQ}(E)$ with $b\approx $3 for MCR with low-momentum ($\approx $ 50-90 MeV/c) and low-excitation ($E\approx $ 10-50 MeV). The PEQ contribution for the first neutron emission depends on the nuclear size, getting smaller as the nuclear size becomes larger. It is estimated to be around $p\approx 0.6 A^{-1/3}$ for the present MCR case by referring to the observed neutron energy spectra \cite{ref7}.

The residual nucleus $^{A-1}_{Z-1}$Y after the first neutron emission de-excites by emitting the second neutron or $\gamma$-rays depending on the excitation energy above or below the neutron threshold energy. The ground state of $^{A-1}_{Z-1}$Y is populated after the $\gamma $ emission. The 
$2$nd neutron n$_2$ is the EQ evaporation neutron, and then the $3$rd neutron is emitted if the residual nucleus after the $2$nd neutron emission is neutron-unbound, and so on. Then, one gets 
finally the residual isotopes of $^{A-x}_{Z-1}$Y with $x$ = 0,1,2,3,... depending on the excitation energy $E$ and the number $x$ of the emitted neutrons. Some of them are $\beta$-unstable RIs.  

\begin{figure}[htb!]
\centering
\includegraphics[scale=0.3]{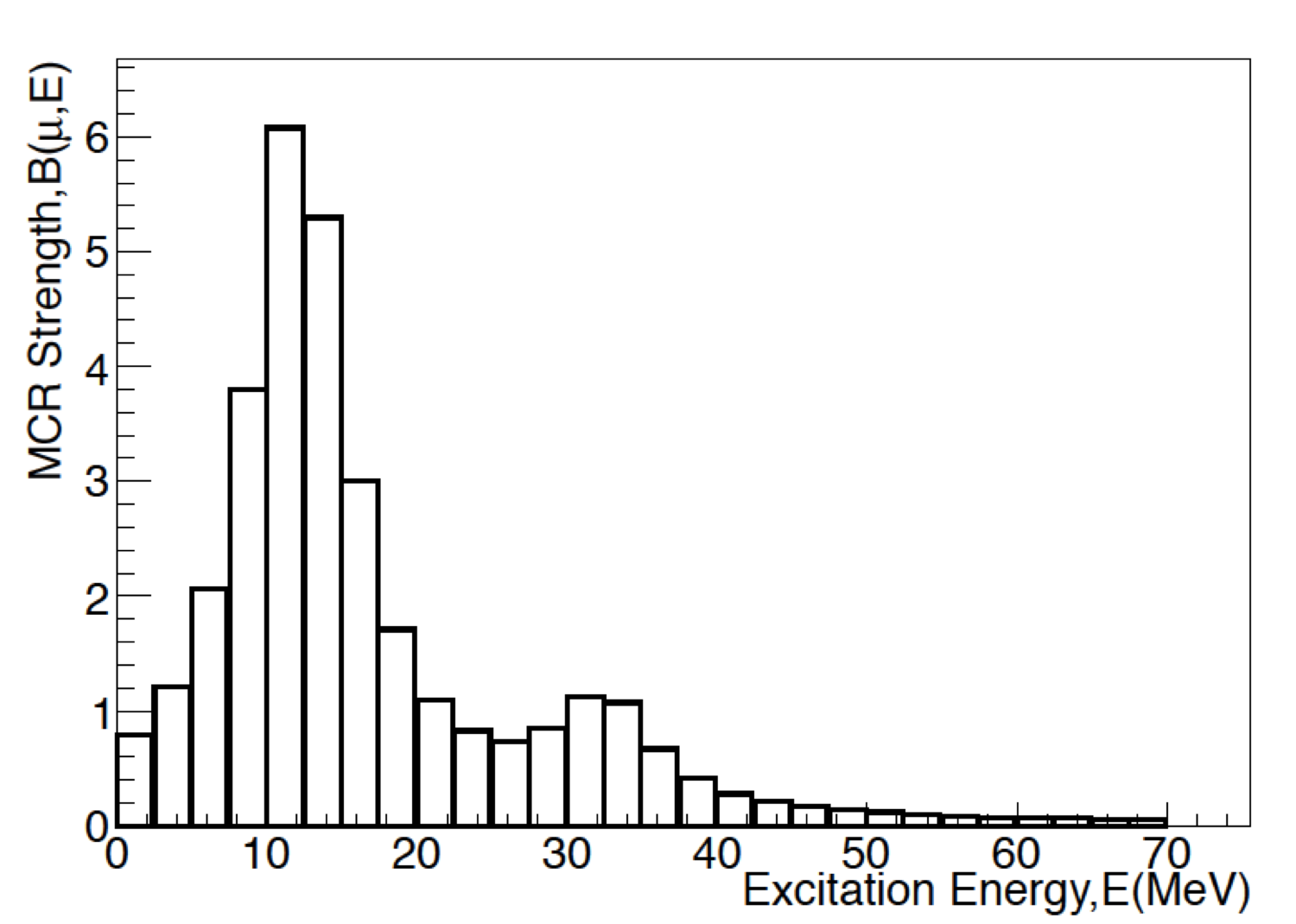}
\caption{\label{figure6} The MCR strength distribution suggested from the experimental RI distribution. $E_{G1}$ and $E_{G2}$ are the MCR GRs at around 12 MeV and 30 MeV.}
\end{figure}

The neutron number $x$ and the mass number $A-x$ distributions reflect the strength distribution $B(\mu,E)$ of the nucleus $^A_{Z-1}$Y$^*$ after MCR, the highly excited states around 30-40 MeV emits 3-4 neutrons while the low excited states around 11-14 MeV emit one neutron as illustrated in Fig.\ref{figure6}. In other words, the GR-like strength around 11-14 MeV leads preferentially to population of $^{A-1}_{Z-1}$Y after 1 neutron emission, and the population of $^{A-x}_{Z-1}$y decrease as $x$ increases. These features are just what have been observed \cite{ref8,ref7,ref12}.

We compare the observed RI mass distribution for MCR on 
$^{100}$Mo with the calculation based on the strength distribution and the EQ/PEQ neutron emission model. The 
obtained RI mass distribution is compared with the observed one in Fig.\ref{figure6}. The agreement with the observed 
data is quite good where $\chi ^2$ is 0.06. The parameters used for the calculation are $E_{G1}$=12 MeV with $\Gamma_1$=8 MeV, $E_{G2}$=30 MeV with $\Gamma_2$=8 MeV, and the cross section 
ratio is $\sigma_1/\sigma_2$=1/6. The first GR corresponds to the large population of the mass $A-1$ with $x$=1 neutron emission, while the second GR match with the population of the RIs with the mass around $A-3$ and $A-4$ with $x$=3-4 neutron emission.

The MCR GR may be compared with the photon capture reaction giant resonance (PCR GR). The energy of 12 MeV is a bit smaller than the PCR GR energy of 14 MeV, but the width of 8 MeV is much larger than the width of 5 MeV for PCR GR. MCR GR consists of  mixed components of $J^{\pi}$=1$^-$, 1$^+$, 2$^-$,... while PCR GR is only one component of $J^{\pi}$=1$^-$.

MCRs in other nuclei have been studies as discussed in the review paper \cite{ref7,ref9,ref10,ref16,ref17}. The one neutron emission is dominant in most MCRs, being consistent with the present observation on $^{100}$Mo, and with the GR strength corresponding to the one neutron emission. Neutron energy spectra were observed for MCRs on $^{32}$S, $^{40}$Ca, $^{207}$Pb and $^{209}$Bi. They reported low-energy EQ and medium energy PEQ components from neutron time-of-flight (TOF) measurement. They are reproduced by the EQ/PEQ neutron emission model with the GR1 and GR2 strength distribution given in equation (6). 

\section{\label{sec:level4}Concluding remarks and Perspectives}

MCR, as the lepton-sector charge exchange reaction (CER) via the weak boson, is shown to be used to study neutrino nuclear responses relevant to DBD and  astro-neutrino reactions. It 
provides unique information on $\beta^+$ side DBD NMEs and astro anti-neutrino NMEs in the energy and momentum regions of $E\approx$5-50 MeV and $p\approx$95-50 MeV/c. 

These are just the regions associated with the neutrino-less DBD and supernova neutrinos. Nb RIs produced by MCR on $^{100}$Mo are identified by measuring delayed $\gamma$-rays characteristic of the RIs. The Nb mass distribution shows large yield at $A$=99 after $x$=1 neutron emission, and decreases gradually as $A$ decreases till $A$=95 ($x$=5). The neutron emission is analyzed in terms of the EQ/PEQ neutron emission model. 

The observed RI distribution reflects the MCR strength distribution with a broad peak at $E\approx$ 11-14 MeV and the small bump at $E\approx$ 30-40 MeV. The broad peak is a kind of $\mu$-capture GR analogous to the E1 photon-capture GR. It is noted that similar GRs are observed for the DBD $\beta^-$-side responses and the astro neutrino responses \cite{ref1,ref4,ref5}.

The RCNP MuSIC DC muon beam  and the J-PARC MLF pulsed muon beam are promising for further studies of neutrino nuclear responses. The lifetime measurement is under progress to study the absolute strength (square of absolute NME). The absolute response, together with the strength distribution, help theories to evaluate the DBD NMEs and astro-neutrino synthesis/interaction NMEs. In other words, nuclear models to be used for DBD NME calculations should reproduce the MCR strength distribution as observed. Experimental studies of MCRs on all DBD nuclei are under progress \cite{ref18}.

The EQ/PEQ neutron emission code was developed for the neutron emission following MCR on the present $^{100}$Mo 
with the large neutron excess and $Z$=42. One has to include proton emissions as well in medium heavy nuclei with less neutron excess since the proton binding energy gets lower relative to the neutron one, and also in light nuclei where the Coulomb barrier gets lower. The EQ/PEQ code with both neutrons and protons are being developed at UTM and RCNP.

The present experiment is made on delayed $\gamma $-rays from RIs produced by MCR to get the yields of the RIs since the 
$\gamma$-ray branching ratios are well known. Then we get the gross structure of the strength distribution up to 
$E\approx$ 60 MeV. The strength (response) to individual states below the neutron threshold energy can be studied by measuring prompt $\gamma$-rays from MCR.

The population of the $i$th state is in principle obtained from the difference between the summed yield of all the $\gamma$-rays from the $i$th state and that of all the $\gamma $-rays feeding the $i$th state from higher states. However, in medium heavy nuclei with rather high level density, it is a challenge to measure accurately all the $\gamma $-rays from and to the individual states.

\end{document}